\documentclass{article}
\usepackage{spconf,amsmath,graphicx,amsfonts,bm}

\usepackage{cite} 

\title{SOURCE CODING OF AUDIO SIGNALS WITH A GENERATIVE MODEL}
\name{Roy~Fejgin$^{1}$ \qquad Janusz~Klejsa$^{2}$ \qquad Lars~Villemoes$^{2}$\qquad Cong~Zhou$^{1}$\qquad}
\address{$^{1}$ Dolby Laboratories, San Francisco, CA, USA\\
$^{2}$ Dolby Sweden AB, Stockholm, Sweden}
\begin{document}
\ninept
\maketitle
\begin{abstract}
We consider source coding of audio signals with the help of a generative model. We use a construction where a waveform is first quantized, yielding a finite bitrate representation. The waveform is then reconstructed by random sampling from a model conditioned on the quantized waveform. The proposed coding scheme is theoretically analyzed. Using SampleRNN as the generative model, we demonstrate that the proposed coding structure provides performance competitive with state-of-the-art source coding tools for specific categories of audio signals. 
\end{abstract}
\section{Introduction}
\label{sec:intro}
We propose a source coding scheme for audio employing a deep generative model that facilitates perceptually-optimized signal quantization with a seamless transition between waveform coding and parametric reconstruction of a coded signal. The scheme is capable of performing bandwidth extension, and of filling the reconstructed spectrum of a signal with plausible structures. In this paper, we provide two examples of scenarios where the proposed scheme outperforms state-of-the-art source coding techniques.
\par Deep generative models have been successfully used for speech coding \cite{Kleijn2018, Klejsa2019, Valin2019, Skoglund2019}, providing a significant improvement to the perceptual quality--bitrate tradeoff. These schemes comprised an encoder computing a parametric (finite bitrate) representation of speech, and a decoder based on a generative model. The speech signal was reconstructed by sampling from a learned probability distribution conditioned on the parametric representation.  
\par Generative models were also used for synthesis of audio signals \cite{Oord2016, Mehri2016, Engel2019}. However, their application to audio coding remains an open problem. An application that is closest to the coding problem is a scheme of the Magenta Project \cite{Hawthorne2018}, where piano waveforms were encoded into MIDI-like representation and then reconstructed from it. This conceptually resembles the mentioned speech coding schemes, where an encoder provides a salient parametric description of the signal to be generated. Perhaps the most obvious disadvantage of such an approach for audio is that the set of salient parameters would depend on signal category (e.g., MIDI-like parametrization would not be suitable to speech). In this work, we address this shortcoming by proposing a coding scheme that uses a generative model conditioned on quantized waveform. Specifically, we consider a source coding setup where a deterministic waveform coder is used to provide a finite-bitrate conditioning for a generative model at the decoder side. 
\par Deep neural networks have already been applied to the audio coding problem in \cite{Porov2018, Shin2019, Deng2019}. However, these schemes are based on discriminative networks. In contrast, generative modeling provides means for synthesis of plausible signal structures. This enhances the perceptual performance at tasks such as bandwidth extension, or noise-fill by creating signal structures that would otherwise be lost due to signal quantization. Signal quantizers capable of providing source matching noise-filling were proposed in \cite{Li2009, Li2010}. However, these schemes were limited to scalar quantizers and used simple probability distributions describing the source. 
\par This paper is organized as follows. In Section \ref{sec:source_coding} we propose a source coding scheme for audio signals and provide its analysis. Then, in Section \ref{sec:coding_with_sRNN} we describe a practical coding scheme based on SampleRNN \cite{Mehri2016}. We evaluate its performance objectively and by means of listening tests in Section \ref{sec:evaluation}.
\vspace{-0.06cm}
\section{Source Coding With A Generative Model}
\label{sec:source_coding}
\begin{figure}[htb]
\centering
\includegraphics[width=\linewidth]{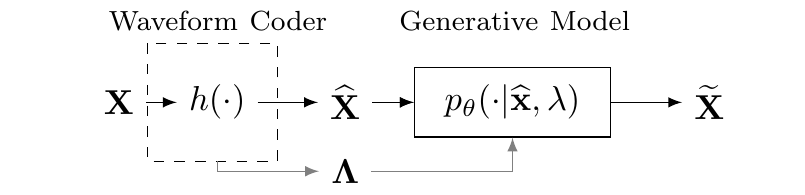} 
\caption{Diagram of the proposed coding scheme. \label{fig:main}}
\end{figure}

We study a coding scheme (shown in Fig. \ref{fig:main}) consisting of a \emph{waveform coder} and a generative model.
\vspace{-0.15cm}
\par We use upper case letters for random variables and lower case letters for their realizations. The waveform coder $h(\cdot)$ operates on signal samples blocked into vectors represented by $\mathbf{X}$, mapping $\mathbf{x}$ to a waveform reconstruction $\widehat{\mathbf{x}}$ and a set of parameters $\bm{\lambda}$, both represented at a finite bitrate. The coder implements a bitrate--distortion tradeoff by using a sample distortion measure---here, perceptually-weighted squared error. The generative model provides a signal reconstruction $\widetilde{\mathbf{x}}$ by random sampling from a conditional probability distribution $p_{\theta}(\cdot|\widehat{\mathbf{x}},\bm{\lambda})$ with parameters $\theta$ trained using a standard negative log-likelihood (NLL) loss.
\par In the following subsection, we analyze an idealized instance of the scheme. We first argue that the scheme tends to preserve the waveform match between $\mathbf{X}$ and $\widetilde{\mathbf{X}}$, which by itself is a useful property. It is then shown that, in the limit of increasing rate, the scheme incurs a performance loss in terms of sample distortion compared to a reconstruction with $\widehat{\mathbf{X}}$. Interestingly, the distortion does not need to increase at low rates. Moreover, an increased sample distortion does not necessarily harm  perceptual performance.  A practical implementation is provided in Section \ref{sec:coding_with_sRNN}, and finally the perceptual benefits are illustrated in Section \ref{sec:evaluation}.

\subsection{Theoretical analysis}
\label{subsec:theory}
We will now discuss how the use of the NLL loss is related 
to the task of waveform coding. Each input signal
in $\mathbf{x} \in \mathbb{R}^d$ is mapped 
to $(\widehat{\mathbf{x}},\bm{\lambda}) = \mathbf{y}  = h(\mathbf{x})$ by the (measurable) deterministic codec 
$h$.  Due to the finite bitrate, the 
image ${\cal{Y}} = h( \mathbb{R}^d ) $ consists of a finite set of points.  
If $\Omega_\mathbf{y} = \{ \mathbf{x} : h(\mathbf{x}) = \mathbf{y} \}$ is the set of signals sharing the codec point $\mathbf{y}$, then $\cup_{\mathbf{y}\in \cal{Y}} \Omega_\mathbf{y}$ is a partition of $\mathbb{R}^d$. 
(In general $\Omega_\mathbf{y}$ can be a complicated high-dimensional object, but we will study a 
toy example in Section \ref{subsec:toy}.) 
Assume $p(\mathbf{x})$ is the probability density of the signal source, and let $p_\mathbf{y}$ be the probability of $\Omega_\mathbf{y}$. Then 
\begin{align}
\label{eq:KL}
\mathbb{E}_{\mathbf{X}\sim p(\cdot)} \{-&\log  p_\theta(\mathbf{X}|h(\mathbf{X}))\} \notag\\
 &=\sum_{\mathbf{y}\in \cal{Y}} p_\mathbf{y} \,\mathbb{E}_{{\mathbf{X}\sim p(\cdot|\mathbf{y})}}\{-\log p_\theta(\mathbf{X}|\mathbf{y})\} \notag \\
&\ge \sum_{\mathbf{y}\in \cal{Y}} p_\mathbf{y} \,\mathbb{E}_{{\mathbf{X}\sim p(\cdot|\mathbf{y})}}\{-\log p(\mathbf{X}|\mathbf{y})\},
\end{align}
where the final inequality follows from the non-negativity of the Kullback-Leibler divergence, $\mathbb{KL}\{p(\cdot|\mathbf{y}) \| p_\theta(\cdot|\mathbf{y}) \} \ge 0$. 

Viewing the left hand side of \eqref{eq:KL} as the idealized NLL loss, this shows that the training will 
encourage the parametric model $p_\theta(\mathbf{x}|\mathbf{y})$ to match the optimal conditional density 
$p(\mathbf{x}|\mathbf{y})$. An important property of this optimal case is that the resulting idealized
scheme of Fig.~\ref{fig:main} will preserve the distribution of the source. Another aspect is that \emph{the noise 
shaping properties of the waveform coder will be inherited} since $p(\mathbf{x}|\mathbf{y})=0$ for $\mathbf{x}\notin\Omega_\mathbf{y}$, which often    
% $p(\mathbf{x}|\mathbf{y}) = \mathbf{1}_{\Omega_\mathbf{y}}(\mathbf{x}) p(\mathbf{x}) / p_\mathbf{y}$, where $\mathbf{1}_{\Omega_\mathbf{y}}$ is the 
%characteristic function of $\Omega_\mathbf{y}$. 
is a perceptually motivated vicinity of the original signal. In practical cases, the tradeoff between minimizing the probability mass outside of $\Omega_\mathbf{y}$ and achieving a good signal source match inside it is of course unknown.

The analysis of squared sample distortion in subspaces and with signal dependent perceptual weighting can be accommodated
by using a positive semi-definite bilinear form depending on $h(\mathbf{x})$.  
If $\mu(\mathbf{z})$ is the expected value of $\mathbf{X}$ in $\Omega_{h(\mathbf{z})}$, one can show, 
by using the partition $\cup_{\mathbf{y}\in \cal{Y}} \Omega_\mathbf{y}$ and the 
linearity of expectation, that 
\begin{multline}
\mathbb{E} \{ \|  \widetilde{\mathbf{X}}-\mathbf{X} \|_{h(\mathbf{X})}^2 \} \\
 =
\mathbb{E} \{ \|  \mathbf{X}-\mu(\mathbf{X}) \|_{h(\mathbf{X})}^2 \} 
+ \mathbb{E} \{ \|  \widetilde{\mathbf{X}}-\mu(\mathbf{X}) \|_{h(\mathbf{X})}^2 \},
\label{eq:meanQ}
\end{multline}
where expectations are with respect to the joint density $p(\widetilde{\mathbf{x}},\mathbf{x}) = p(\widetilde{\mathbf{x}}|\mathbf{x})p(\mathbf{x}) =  p_\theta(\widetilde{\mathbf{x}}|h(\mathbf{x}))p(\mathbf{x})$ of the coding scheme.

The first term on the right hand side of \eqref{eq:meanQ} is a (well-known) lower bound on the distortion, achieved by taking $\mu(\mathbf{X})$ as the reconstruction. For the NLL-optimal scheme 
$p_\theta(\cdot|h(\mathbf{x})) = p(\cdot|h(\mathbf{x}))$, the two terms of the right hand side of 
\eqref{eq:meanQ} are equal, 
%(since $h(\widetilde{\mathbf{X}}) = h(\mathbf{X})$, $\mu(\widetilde{\mathbf{X}}) = \mu(\mathbf{X})$, and $\widetilde{\mathbf{X}}$ has the same distribution as $\mathbf{X}$), 
resulting in a $3$~dB loss of 
performance compared to $\mu(\mathbf{X})$,
\begin{equation}
\mathbb{E} \{ \|  \widetilde{\mathbf{X}}-\mathbf{X} \|_{h(\mathbf{X})}^2 \} 
= 2\,\mathbb{E} \{ \|  \mathbf{X}-\mu(\mathbf{X}) \|_{h(\mathbf{X})}^2 \}. 
\label{eq:3dBwall}
\end{equation}
Obviously, this loss also holds relative to a deterministic decoder with $\widehat{\mathbf{x}}\approx\mu(\mathbf{x})$, which is more likely to happen in a high rate limit characterized by a flat signal 
distribution in $\Omega_\mathbf{y}$.      

\subsection{Toy example}
\label{subsec:toy}
To illustrate the concepts, we will study a synthetic example where the generative model outperforms the deterministic decoder. Hence, we are not in the high rate case outlined at the end of Section \ref{subsec:theory}.

Let the signal source be random sines of unit amplitude,
$ x[k] = \cos\left[ \pi (z_1 k + 2 z_2) \right] $, $k=1,\dots,d$, with $z_1, z_2$ i.i.d. 
uniformly distributed on $[0,1]$, and let the waveform coder use scalar quantization 
with step size $\Delta > 0$ with midpoint reconstruction. This means that 
$\widehat{x}[k] =  \Delta\operatorname{round}(x[k]/\Delta)$, and 
the sets $\Omega_y$ are Voronoi cells of the quantizer which in this case are 
hypercubes of side length $\Delta$ in 
$\mathbb{R}^d$. For $d = 10$, it is feasible to implement $p(\mathbf{x}|\mathbf{y})$ by drawing random vector samples of $\mathbf{x}$ 
until $\mathbf{x}\in \Omega_\mathbf{y}$. We estimated the left hand side of 
\eqref{eq:meanQ}, (without $h(\mathbf{X})$), by $10000$ trials. For the 
generative model case we also created the mean over $10$ realizations from $p(\mathbf{x}|\mathbf{y})$. The 
results normalized to per sample distortion are given in Table \ref{tab:toy}.
\vspace{-0.3cm}
\begin{table}[h!]
\centering
\caption{\label{tab:toy} Mean squared distortions for the toy example.}
\begin{tabular}{c|c|c|c}
\hline
 $\Delta$   & \text{Midpoint} & \text{Sampling} & \text{Mean of 10 realizations} \\ 
 \hline
 0.5  & 0.026 & 0.011 & 0.0056 \\
 1   & 0.11 & 0.068 & 0.038  \\
 %delta = .5  d = 10 % 10000 monte carlo
%0.0261    0.0470    0.0105    0.0056
%15.8262   13.2826   19.7997   22.4837 dB instead ?
%0.0262    0.0464    0.0102    0.0056
%15.8223   13.3388   19.9158   22.533
%delta = 1  d = 10 % 10000 monte carlo
%0.1134    0.1971    0.0695    0.0380
%0.1134    0.1958    0.0683    0.0382
 \hline
\end{tabular}
\end{table}

As it can be seen, powerful signal source modeling leads to a significant 
advantage over the scalar midpoint quantizer. A comparison with \eqref{eq:3dBwall} 
also reveals that the mean value of $10$ 
samples approaches the optimum. 
However, one should keep in mind that among the three systems, 
only the middle one preserves the source distribution. 
\vspace{-0.35cm}
\section{Source Coding with SampleRNN}
\label{sec:coding_with_sRNN}
Next, we describe a practical implementation of the scheme discussed in Section~\ref{sec:source_coding}. The scheme comprises a waveform codec operating in an MDCT domain and a generative model based on the SampleRNN \cite{Mehri2016}, which is conditioned by waveform reconstructions obtained from the waveform codec.

\subsection{Waveform coder}
\label{sec:waveform}
\begin{figure}[htb]
\centering
\includegraphics[width=\linewidth]{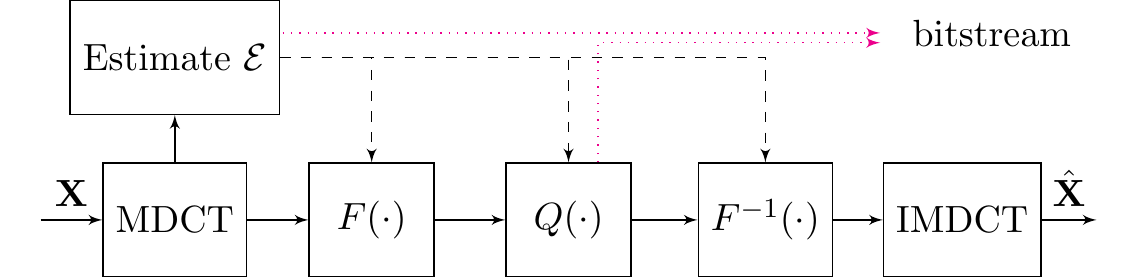}
\caption{Diagram of the waveform coder providing conditioning. \label{fig:tcs}}
\end{figure}

In this work we use a simple waveform coder operating in the MDCT domain, which is shown in Fig.~\ref{fig:tcs}. The input signal $X$ is framed to facilitate application of an MDCT with a stride of $320$ samples (at sampling frequency of $f_s=16$ kHz). The coefficients of the transform are blocked into $N$ non-uniform, non-overlaping frequency bands. For an $n$-th band, a variance of the coefficients is computed and quantized with a $3$~dB step, yielding an index $i_{\text{env}}(n)$. The quantized values are blocked into a vector $\mathcal{E}$ and coded into a bitstream using frequency differential coding with a Huffman codebook.
\par On the encoder side, the MDCT coefficients are first spectrally flattened by $F(\cdot)$ according to the envelope $\mathcal{E}$. The flattened MDCT lines are then quantized by a set of quantizers selected to fulfil a per-frame bitrate constraint. The set of quantizers $[m_0, \dots, m_M]$ is ordered providing incremental increases of SNR by $1.5$~dB between each $m_n$ and $m_{n+1}$. Each $m_n$ is associated with a Huffman codebook.
\par For every coded block, the rate allocation process is constrained by the total number of bits allocated to that block. It is controlled by $m_n = i_{\text{env}}(n) - i_{\text{offset}}$, where $i_{\text{offset}}$ is an integer common to all the frequency bands and $m_n$ is limited so that $0\leq m_n\leq M$. The value of $i_{\text{offset}}$ is determined by a binary search, which resembles the reverse water-filling procedure in a perceptually weighted domain. The perceptual effect of this rate allocation is that the SNR within a frame is allocated proportionally to the square root of the spectral envelope (allocating $1.5$~dB SNR increase for every increase of the in-band envelope value by $3$~dB).
\par On the decoder side, the MDCT lines are reconstructed in the flattened domain, and then the inverse spectral flattening $F^{-1}(\cdot)$ is applied. The inverse flattening is controlled by $\mathcal{E}$, which is decoded from the bitstream along with quantized transform coefficients and the rate allocation parameter $i_{\text{offset}}$.
\vspace{-0.25cm}
\subsection{Conditional SampleRNN}
SampleRNN is a deep neural generative model proposed in \cite{Mehri2016} for generating raw audio signals by sampling them from a trained model $p_{\theta}(x)$. It consists of a series of multi-rate recurrent layers which are capable of modeling the dynamics of a sequence at different time scales and a multilayer perceptron (MLP) allowing usage of parameterized simple distributions. \par SampleRNN models the probability of a sequence of audio samples blocked in $\mathbf{x}$ by factorizing the joint distribution into the product of the scalar sample distributions conditioned on all previous samples. This facilitates an efficient implementation, where a single scalar sample is drawn at a time. Here we use a conditioned model $p_{\theta}(\mathbf{x}|\mathbf{y})$. % Let us denote the scalar samples blocked in $\mathbf{x}$ by $[s_0, \dots, s_n]$. 
The model operates recursively according to
\begin{equation}
p_{\theta}(\mathbf{x}|\mathbf{y})=\prod_{i=1}^{T}p(x_i|x_1,\dots,x_{i-1}, \mathbf{y}).
\label{eq:pxh}
\end{equation}  
We use a model similar to the one described in \cite{Klejsa2019}. It is a four-tier SampleRNN with the conditioning provided to each tier through convolutional layers. We denote the frame size used by the $k$-th tier $\text{TS}^{(k)}$ ($1\leq k\leq 4$) and denote the number of logistic mixture components $L$. The values of these hyperparameters are specified in section \ref{sec:evaluation}. The output layer utilizes the discretized mix of logistics technique \cite{oord2017parallel} to generate 16-bit outputs. The differences to the model from \cite{Klejsa2019} are as follows. The model here is conditioned on $\mathbf{y}$ comprising frames of signal domain samples $\widehat{\mathbf{x}}$ reconstructed by the waveform codec and the associated values of the quantized signal envelope in $\mathcal{E}$ (corresponding to $\bm{\lambda}$ in Fig.~\ref{fig:main}). The model operates with a look-ahead which improves the performance. This is done by processing the conditioning vector with $3\times1$ convolution layer, which results in a lookahead of two codec frames. Another difference to the model from \cite{Klejsa2019} is an update to the MLP operation, where the MLP block, in addition to the conditioning described above, has access to the coded waveform processed through a convolutional layer utilizing a $319\times1$ kernel centered on the coded sample aligned with the prediction target. We use NLL as the training objective. We train the model using ADAM \cite{kingma2014adam} with a learning rate of $\text{2e-4}$ and reduce the learning by a factor of 0.3 when the validation loss stops improving. In generation, we perform random sampling from the model. 
\vspace{-0.1cm}
\section{Experiments}
\label{sec:evaluation}
In this section we provide results of subjective evaluation of the proposed source coding scheme in two coding tasks. The first task comprised coding of piano excerpts. The second task comprised coding of speech. In each of the tasks we compare the scheme against state-of-the-art codecs that are meant to represent source coding tools that would be typically used for the specific signal category considered in a coding task. We also provide examples and measurements corroborating the theory outlined in Section~\ref{sec:source_coding}.
\subsection{Subjective evaluation}
In the first experiment we evaluate the proposed source scheme in a piano coding task. We trained the generative model using the Maestro dataset \cite{Hawthorne2018}, which was divided into non-overlapping training, validation and test sets. In order to provide the conditioning we used the waveform coder described in Section~\ref{sec:waveform}. SampleRNN (``sRNN'') was configured with $\text{TS}^{(1)}=8$, $\text{TS}^{(2)}=8$, $\text{TS}^{(3)}= 64$, $\text{TS}^{(4)}= 320$ and $L=1$. We conducted a MUSHRA-like listening test\cite{bs.1534-3} on the test set items, where we compared the proposed scheme against Opus\cite{Valin2012} and AAC (which is a core of state-of-the-art audio codecs \cite{Herre2008, Quackenbush2013}), and the baseline waveform (``Waveform'') coder operating at 16~kb/s. The conditions also included a hidden reference (16~kHz sampling) and a $3.5$~kHz low pass anchor.
\begin{figure}[htb]
\centering
\includegraphics[width=\linewidth]{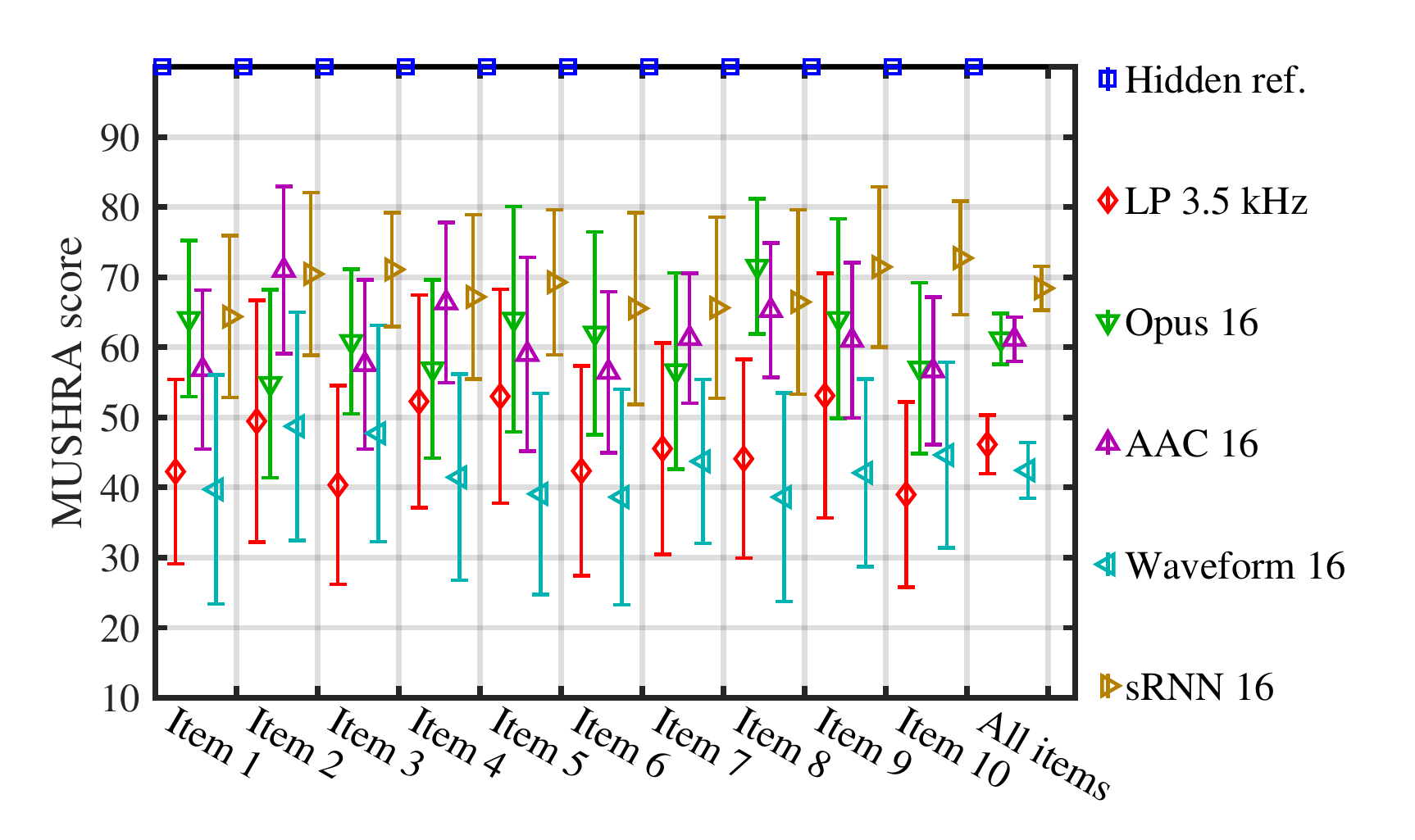}
\caption{Listening test results for Maestro piano (11 listeners, 95\% confidence intervals, Student's t-distribution). \label{fig:piano}}
\end{figure}
\vspace{-0.5cm}
\begin{figure}[htb]
\centering
\includegraphics[width=\linewidth]{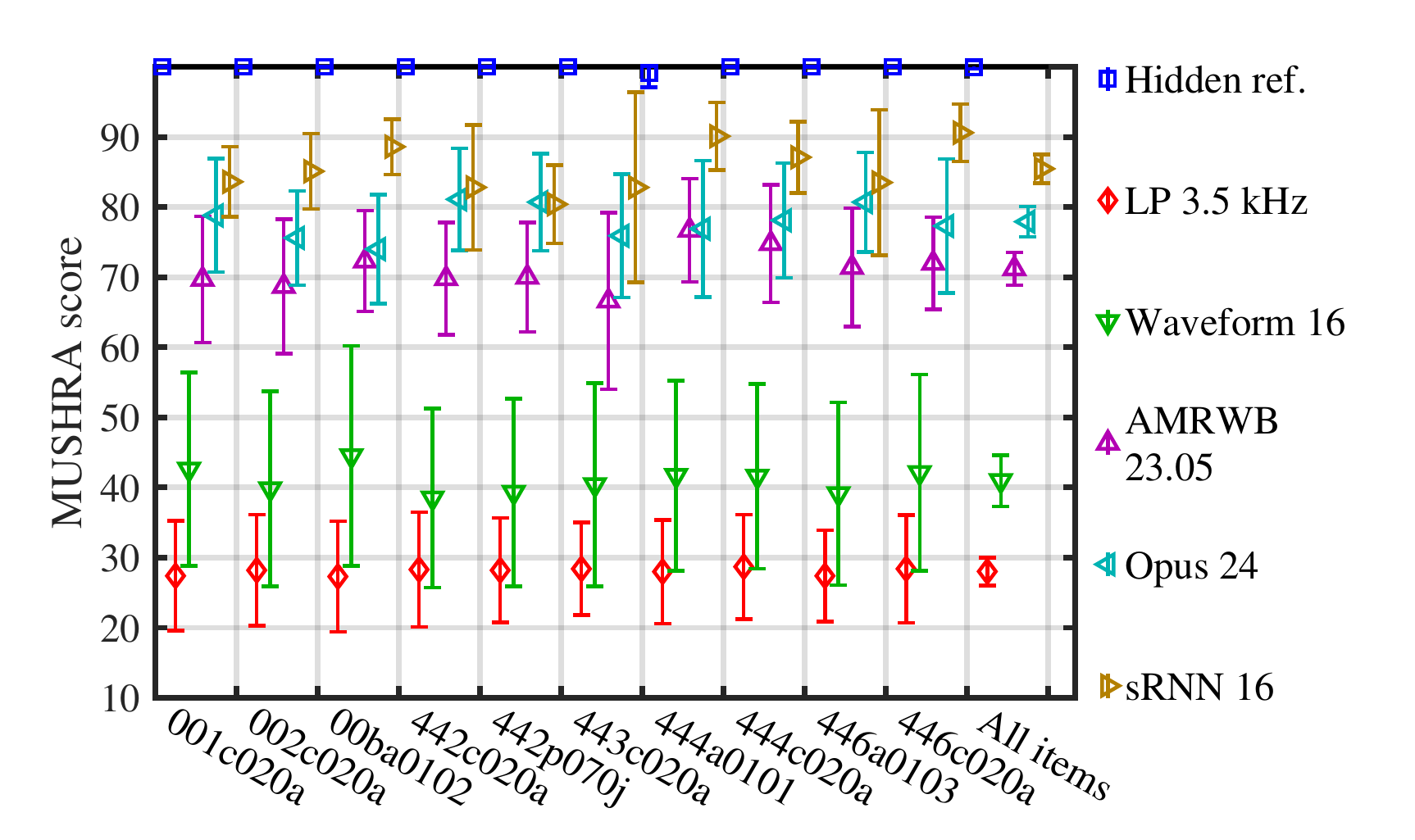}
\caption{Listening test results for WSJ0 speech (10 listeners, 95\% confidence intervals, Student's t-distribution). \label{fig:speech}}
\end{figure}
\begin{figure*}[ht!]
\vspace{-0.55cm}
\begin{minipage}{.33\textwidth} 
  \includegraphics[width=.99\textwidth]{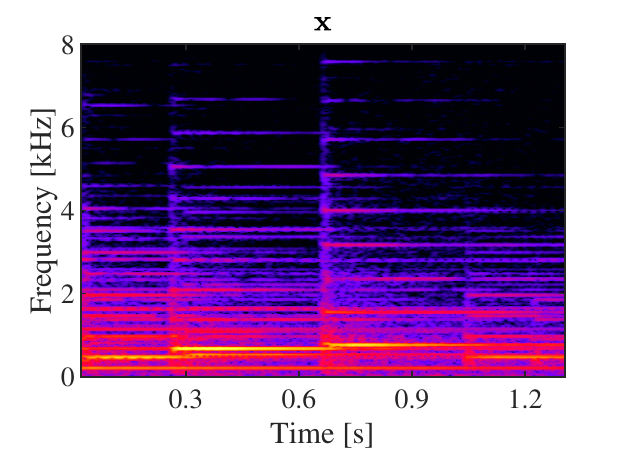}\vspace{-0.3cm}\hfill
\end{minipage}  
\begin{minipage}{.33\textwidth}  
  \includegraphics[width=.99\textwidth]{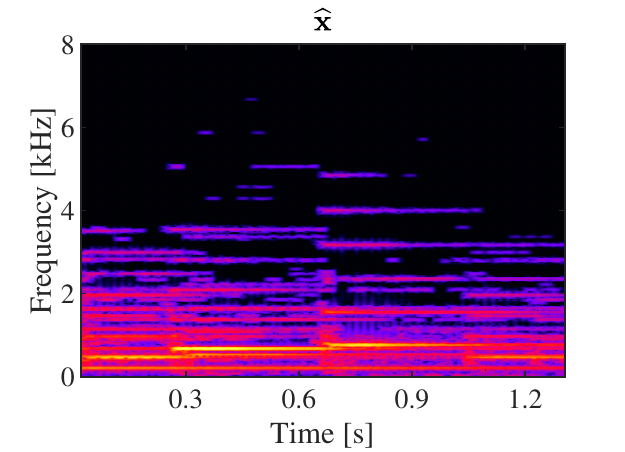}\vspace{-0.3cm}\hfill
\end{minipage}  
\begin{minipage}{.33\textwidth}   
  \includegraphics[width=.99\textwidth]{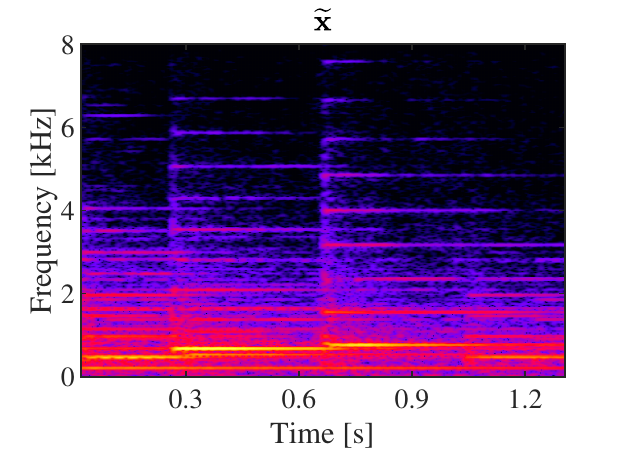}\vspace{-0.3cm}
\end{minipage}
  \caption{Spectrograms: (left) reference $\mathbf{x}$; (center) reconstruction from waveform codec $\widehat{\mathbf{x}}$; (right) reconstruction of the proposed scheme $\widetilde{\mathbf{x}}$. \label{fig:spectrograms}}
\end{figure*}

\begin{figure*}[ht!]
\begin{minipage}{.33\textwidth} 
  \includegraphics[width=.99\textwidth]{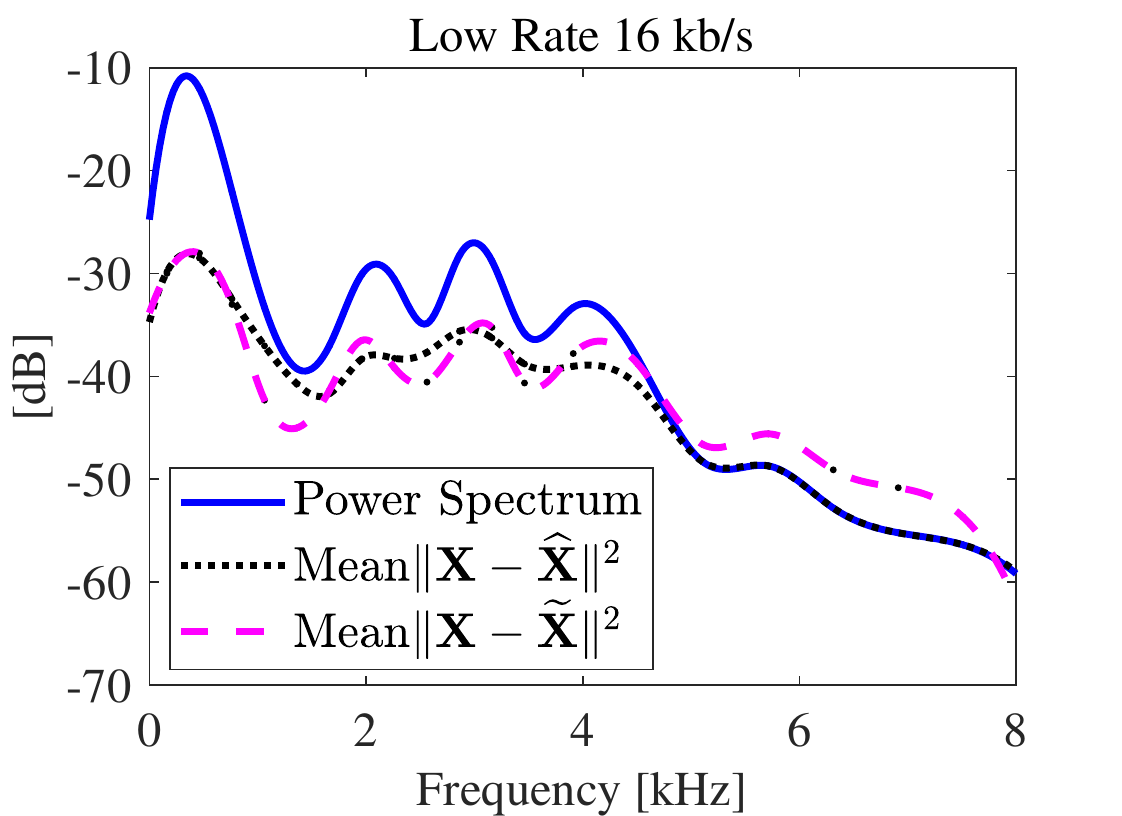}\vspace{-0.35cm}\hfill
  (a)
\end{minipage}  
\begin{minipage}{.33\textwidth}  
  \includegraphics[width=.99\textwidth]{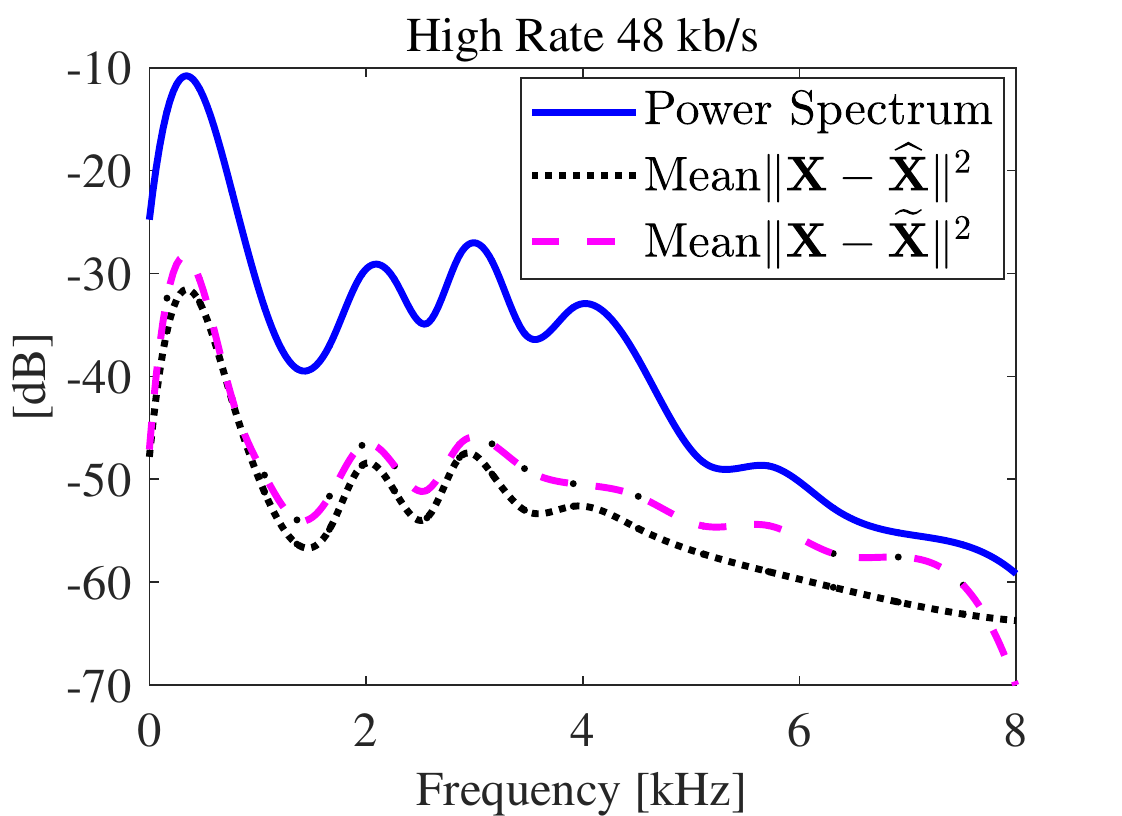}\vspace{-0.35cm}\hfill
  (b)
\end{minipage}  
\begin{minipage}{.33\textwidth}   
  \includegraphics[width=.99\textwidth]{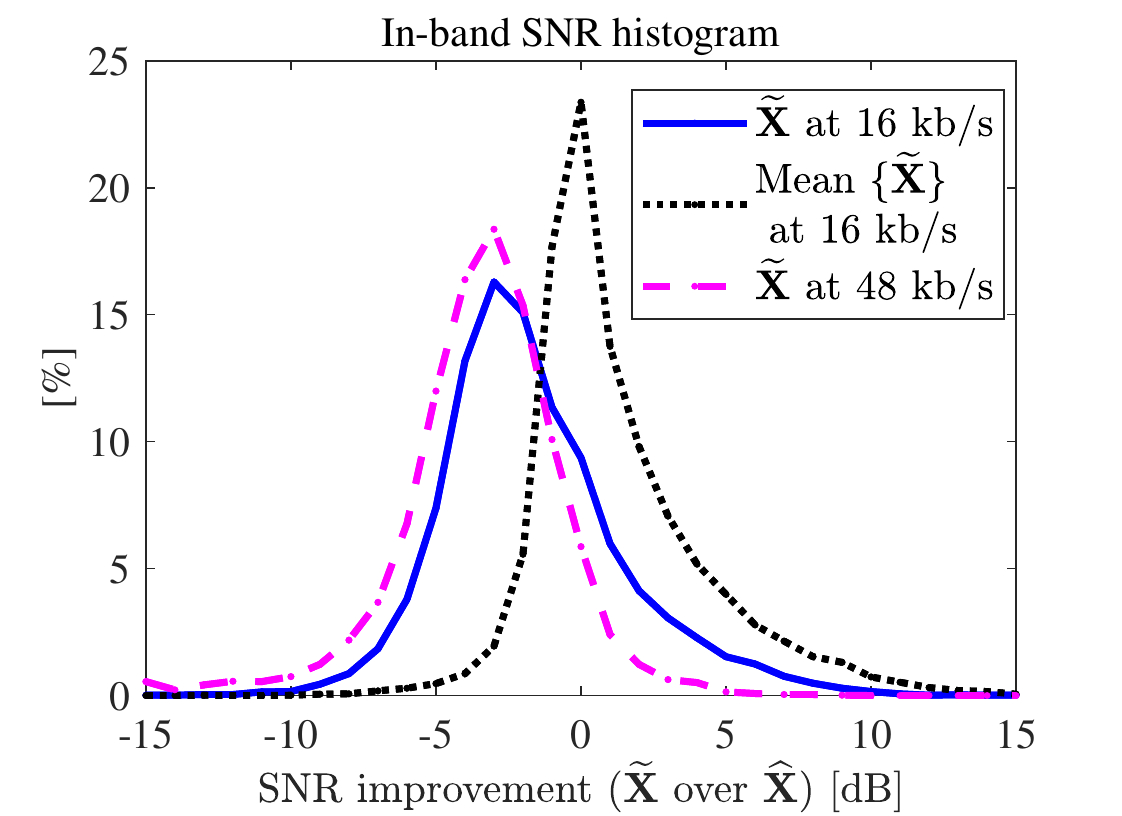}\vspace{-0.35cm}\hfill
  (c)
\end{minipage}
  \caption{Examples of a power spectrum of a signal segment and the corresponding power spectrum of sample distortion for (a)~low rates and (b)~high rates; (c) histograms of SNR improvement in bands of the waveform codec with respect to the waveform codec reconstruction. \label{fig:main_fig}}
\end{figure*}
The results of the first listening test are shown in Fig.~\ref{fig:piano}. It can be seen that the proposed scheme significantly outperforms the baseline waveform coder, but also remains competitive to AAC. The advantage of AAC compared to Opus and the baseline waveform codec is due to its usage of window-switching, which facilitates usage of long MDCT transforms (64~ms stride) where it is perceptually advantageous. 
\par In the second experiment, we evaluated the proposed scheme in a speech coding task. The results are shown in Fig.~\ref{fig:speech}. In this case, we trained the generative model using the WSJ0 dataset \cite{wsj0} which was divided into training, validation and test sets with non-overlapping speakers. SampleRNN was configured with $\text{TS}^{(1)}=2$, $\text{TS}^{(2)}=2$, $\text{TS}^{(3)}= 16$, $\text{TS}^{(4)}= 160$ and $L=10$. We performed a similar subjective test to the one in the previous experiment. This time the conditions included AMR-WB\cite{Bessette2002} at 23.05~kb/s (as it is a commonly included anchor in evaluation of speech codecs based on generative models), Opus at 24~kb/s; and also the baseline waveform codec and the proposed source coding scheme, both operating at 16~kb/s. It can be seen that the proposed scheme outperforms the waveform baseline by a large margin and that it remains competitive with the conditions representing state-of-the-art.
\par The significant perceptual advantage of the proposed scheme over the waveform baseline becomes apparent while inspecting spectrograms of the reconstructed signals. For example, in Fig.~\ref{fig:spectrograms} we illustrate this for a signal from the piano coding experiment.
\vspace{-0.35cm}
\subsection{Objective measurements}
An interesting property of the proposed scheme is that it allows some degree of control of spectral shaping of sample distortion even though the reconstructed signals are generated by random sampling. For example, the  scheme described in Section~\ref{sec:coding_with_sRNN} uses a waveform codec with a perceptual rule allocating the distortion proportionally to the square root of the frequency envelope of the signal. In Fig.~\ref{fig:main_fig}a, we show an example of the power spectrum of a signal segment plotted along with the error spectrum of $\widehat{\mathbf{X}}$ and the error spectrum of $\widetilde{\mathbf{X}}$. It can be seen that the proposed source coding scheme closely follows the error shaping of the waveform codec. In the provided example, in the mid-frequency range the reconstruction with $\widetilde{\mathbf{X}}$ had lower error. At high bitrates (see Fig.~\ref{fig:main_fig}b) the noise shaping of the baseline waveform codec is still followed, but the average performance gap in terms of squared error grows to 3~dB when compared to the baseline waveform codec.
\par In order to provide an overview, we performed SNR measurements within the frequency bands of the baseline waveform codec and we compared the synthesis from SampleRNN to the baseline. The results were plotted as a histogram of SNR improvement over the waveform baseline in Fig.~\ref{fig:main_fig}c. It can be seen that the histograms for the low rate case and for the high rate case are concentrated around -3~dB. The histogram for the 16~kb/s version is more skewed towards the positive improvements likely due to the suboptimality of the baseline waveform codec at low rates. We note that while the scheme on average performs worse in terms of the SNR than the baseline, it provides a far superior perceptual performance. 
\par The proposed interpretation of the scheme has predictive power. For example, one can expect that squared error performance could be improved by taking 
$\widetilde{\bm{\mu}} = \mathbb{E}_{\widetilde{\mathbf{X}}\sim p_{\theta}(\cdot|\mathbf{y})}\{\widetilde{\mathbf{X}}\}$ as the reconstruction (if $p(\cdot)$ is approximated well by $p_{\theta}(\cdot)$). Since computing the expectation directly is difficult, instead we approximated $\widetilde{\bm{\mu}}$ by averaging 10 realizations of $\widetilde{\mathbf{X}}$. The result is shown in Fig.~\ref{fig:main_fig}c, where it can be seen that not only was the $3$~dB gap closed, but also the squared error performance improved over the baseline codec. This corroborates our theoretical interpretation of the scheme. We note that while such averaging can lead to an SNR improvement, this does not necessarily imply a perceptual improvement. In practice there is a non-trivial tradeoff between preservation of a waveform match and preservation of the distribution of the signal.
\vspace{-0.1cm}
\section{Conclusion}
\label{sec:conclusion}
\vspace{-0.17cm}
We proposed a source coding scheme based on a generative model that combines advantages of waveform coding and parametric coding in a seamless manner. When trained for a signal category, the scheme outperforms state-of-the-art source coding techniques. Moreover, the coding scheme can be used together with a perceptual model for allocating the coding distortion. The operation of the scheme and its performance can be described and predicted analytically.
\vspace{-0.35cm}
\section{Acknowledgment}
\vspace{-0.2cm}
The authors would like to thank Jordi Pons, Mark Vinton, Per Hedelin and Heiko Purnhagen for useful discussions and Richard Graff for help with listening tests.
% Below is an example of how to insert images. Delete the ``\vspace'' line,
% uncomment the preceding line ``\centerline...'' and replace ``imageX.ps''
% with a suitable PostScript file name.
% -------------------------------------------------------------------------
%\begin{figure}[htb]
%
%\begin{minipage}[b]{1.0\linewidth}
%  \centering
%  \centerline{\includegraphics[width=8.5cm]{image1}}
%%  \vspace{2.0cm}
%  \centerline{(a) Result 1}\medskip
%\end{minipage}
%%
%\begin{minipage}[b]{.48\linewidth}
%  \centering
%  \centerline{\includegraphics[width=4.0cm]{image3}}
%%  \vspace{1.5cm}
%  \centerline{(b) Results 3}\medskip
%\end{minipage}
%\hfill
%\begin{minipage}[b]{0.48\linewidth}
%  \centering
%  \centerline{\includegraphics[width=4.0cm]{image4}}
%%  \vspace{1.5cm}
%  \centerline{(c) Result 4}\medskip
%\end{minipage}
%%
%\caption{Example of placing a figure with experimental results.}
%\label{fig:res}
%%
%\end{figure}

%\vfill
%\pagebreak

 
\bibliographystyle{IEEEbib}
\begin{small}
\bibliography{strings,refs}
\end{small}
\end{document}